\documentclass[doublecol]{epl2}

\title{Doping dependence of the upper critical field in La$_{2-x}$Sr$_x$CuO$_4$ from specific heat}
\shorttitle{Doping dependence of the upper critical field in La$_{2-x}$Sr$_x$CuO$_4$ from specific heat}

\author{Yue Wang \and Hai-Hu Wen\thanks{E-mail: \email{hhwen@aphy.iphy.ac.cn}}}
\shortauthor{Y. Wang and H.-H. Wen}

\institute{National Laboratory for Superconductivity, Institute of
Physics and Beijing National Laboratory for Condensed Matter
Physics, Chinese Academy of Sciences, P. O. Box 603, Beijing 100080,
People's Republic of China}


\abstract{The low-temperature specific heat of
La$_{2-x}$Sr$_x$CuO$_4$ (LSCO) single crystals in magnetic field $H$
up to 12~T has been examined over a wide range of doping ($0.063\leq
p \leq0.238$). From this we have mapped the upper critical field
$H_{c2}$ of LSCO across the entire superconducting diagram. It is
found that the $H_{c2}$ shows a doping dependence similar to that of
the critical temperature $T_c$. We have discussed the implications
of the result and proposed that there may be an effective
superconducting energy scale responsible for the $H_{c2}$ behavior
in the underdoped region.}

\begin{document}

\maketitle

\section{Introduction}

Determining the fundamental parameters of the superconducting state
in high-$T_c$ cuprates is crucial to understanding the nature of the
high-temperature superconductivity. The upper critical field
$H_{c2}$ is one such quantity which is directly correlated with the
microscopic coherence length $\xi$. Convenient methods for
determining $H_{c2}$ mainly come from the resistive transport or
magnetization measurements. For conventional low-$T_c$
superconductors, their low or moderate $H_{c2}$ enables one to draw
the line of $H_{c2}(T)$ in the field-temperature phase diagram from
the critical temperature $T_c$ to $T\rightarrow0$~K with the
application of the magnetic field $H$ to suppress the
superconductivity. Hence the zero-temperature $H_{c2}(0)$ can be
accurately accessed. Moreover, it is shown that the temperature
dependence of $H_{c2}(T)$ in these systems can be well described by
the Werthamer-Helfand-Hohenberg (WHH) theory~\cite{Werthamer66}.
According to this theory, the $H_{c2}(0)$ can be estimated by the
slope d$H_{c2}/$d$T$ in the vicinity of $T_c$. Therefore in many
cases through investigating the behavior of $H_{c2}(T)$ near $T_c$
one can also satisfactorily obtain the $H_{c2}(0)$ of the sample.

In contrast, for high-$T_c$ cuprates the $H_{c2}(0)$ is inherently
huge in parallel with their high $T_c$. In most cases the
superconductivity could be removed only with very intense magnetic
field $H$, which is not always accessible in
experiments~\cite{Ando99}. Thus most determination of $H_{c2}(0)$ in
high-$T_c$ cuprates relies on the extrapolation of the
high-temperature $H_{c2}(T)$ to $T\rightarrow0$~K based on the WHH
theory. However, recent considerable resistive transport
measurements indicate that the mapped $H_{c2}(T)$ shows unusual low
temperature upward curvature, which is inconsistent with the
saturation predicted by the WHH theory~\cite{Vedeneev06}. This
observation casts doubt on estimating $H_{c2}(0)$ using such
extrapolation procedure.

Presumably the above difficulty is thought to arise from the strong
superconducting fluctuations in high-$T_c$ cuprates, especially for
the underdoped region~\cite{Bonn06}. The recent Nernst effect and
torque magnetometry experiments highlighted this
proposal~\cite{Wang06,Wang05}. It is found that the vortex Nernst
signal or diamagnetic magnetization persists well above $T_c$,
indicating the survival of superconducting correlations at high $T$
though the coherent superconductivity has disappeared at $T_c$. By
tracing the field scale at which the Nernst or diamagnetic signal
vanishes, the $H_{c2}$ has been defined which is higher than that
determined in resistive transports and importantly does not become
zero through the $T_c$.

In the $H$-$T$ phase diagram, the $H_{c2}(T)$ line signifies the
transition from the mixed state to the normal state. From the
viewpoint of the specific heat (SH), associated with this transition
is the increase of the electronic density of states (DOS) with
increasing $H$ and eventually the recovery of the normal state DOS
at $H=H_{c2}$. Moreover, theory has suggested a robust connection
between the DOS in the mixed state and that in the normal
state~\cite{Hussey02}. Thus this recognition actually implies that
there is a well-defined way to evaluate the $H_{c2}(0)$ from
SH~\cite{Gao05}. In this letter we present an analysis of the
low-temperature SH in a series of La$_{2-x}$Sr$_x$CuO$_4$ (LSCO)
single crystals. The field-induced increase of the electronic SH in
the mixed state has been quantitatively determined. Combination with
the normal state electronic SH available in the literature, the
$H_{c2}(0)$ has been extracted in a wide doping range across the
whole superconducting phase diagram. It is found that the
$H_{c2}(0)$ becomes larger as one moves from the underdoped region
to the optimal doping point, and then falls with increasing doping
in the overdoped region, forming a ``dome" shape like the $T_c$. We
have discussed the implications of this finding and proposed that
there is an effective superconducting energy scale responsible for
the $H_{c2}(0)$ behavior in the underdoped region.

\section{Experiment}

\begin{table*}
\caption{Table I. Doping dependence of the $H_{c2}$ for LSCO
obtained from specific heat. The units of $T_c$, $A$, and $H_{c2}$
are K, mJ~mol$^{-1}$~K$^{-2}$~T$^{-0.5}$, and T respectively. The
data for $x=0.19$ are quoted from Ref.~\cite{Nohara00}.}
\label{tab.1}
\begin{center}
\begin{tabular}{ccccccccccccc}
 \hline
$x$           & 0.063  & 0.069  & 0.075  & 0.09  & 0.11  & 0.15  & 0.178  & 0.19  & 0.202  & 0.218  & 0.22  & 0.238  \\
 \hline
$T_c$         & 9.0    & 12.0   & 15.7   & 24.6  & 29.3  & 36.1  & 36.0   & 32.0  & 30.5   & 25.0   & 27.4  & 20.0 \\
$A$           & 0.26   & 0.28   & 0.26   & 0.28  & 0.32  & 0.57  & 0.94   & 1.2   & 1.33   & 1.55   & 1.8   & 2.37   \\
$H_{c2}$      & 16     & 17     & 39     & 53    & 75    & 82    & 56     & 45    & 45     & 44     & 36    & 17     \\
 \hline
\end{tabular}
\end{center}
\end{table*}

We have performed the low-temperature SH measurement in LSCO
throughout the entire superconducting phase
diagram~\cite{Wen05,Wang07,Wen04}. The samples used for the study
were single crystals prepared by the traveling-solvent floating-zone
method. The superconducting transition temperatures $T_c$, defined
as the onset of the diamagnetic signal in the magnetic
susceptibility, are summarized in table \ref{tab.1}. The hole doping
level of the sample $p$ is simply regarded as the Sr concentration
$x$, which ranges from 0.063 to 0.238. The low-temperature SH was
carried out on an Oxford Maglab cryogenic system with a thermal
relaxation technique~\cite{Wen04,Liu05}. The data below 12~K were
analyzed for magnetic field $H$ up to 12~T with $H$ parallel to the
$c$ axis of the sample.

\section{Results and discussion}

\begin{figure}
\onefigure{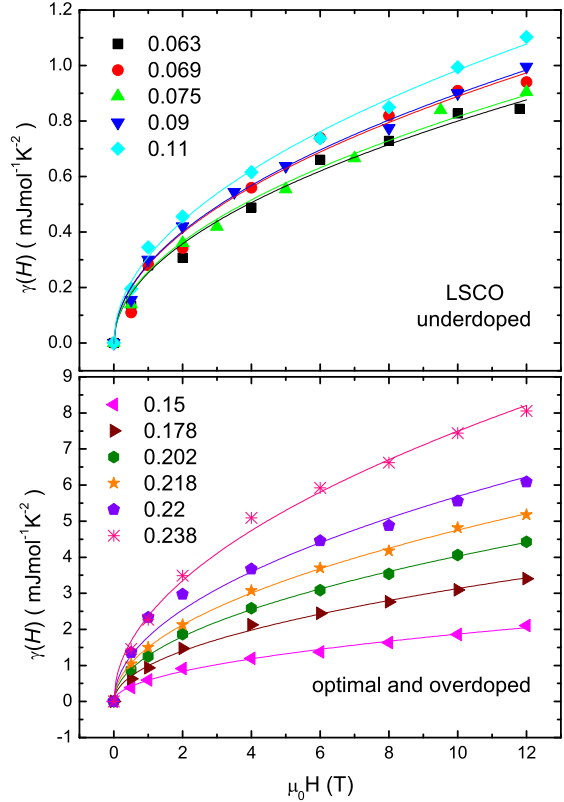} \caption{Field dependence of the increase in
the electronic SH, $\gamma(H)$, at $T\rightarrow0$ K for all samples
(symbols). Note that for $x=0.238$ the $\gamma(H)$ has been
corrected with the estimated superconducting volume fraction of the
sample~\cite{Wang07}. The solid curves are fits to the data with
$\gamma(H)=A\sqrt{H}$.} \label{fig.1}
\end{figure}

From the raw data, we have reliably separated the electronic SH
$C_{\mathrm{el}}=\gamma T$ from other contributions such as the
phonon SH and the possible Schottky anomaly~\cite{Wen04,Wang07}. In
magnetic field $H$, for a superconductor, there is an increase of
the electronic SH in the mixed state, denoted as $\gamma(H)T$. To
investigate the field dependence of the $\gamma(H)$ of the sample is
a useful method to identify the symmetry of the superconducting gap.
For a conventional $s$-wave superconductor, in the mixed state the
electronic DOS comes mainly from the vortex core regions. Since the
vortex number increases linearly as $H$ increases, the
$C_{\mathrm{el}}$ and hence the $\gamma(H)$ is proportional to $H$.
For a $d$-wave superconductor, however, Volovik first pointed out
that the electronic DOS is actually dominated by contributions from
the outer regions of the vortex in a magnetic
field~\cite{Volovik93}. While the vortex number increases linearly
with $H$, the inter-vortex distance is inversely proportional to
$\sqrt{H}$. Thus both effects result in the $C_{\mathrm{el}}$ of a
$d$-wave superconductor showing a $\sqrt{H}$ dependence. In
experiments, the $\sqrt{H}$ behavior of the $C_{\mathrm{el}}$, that
is, $\gamma(H)\propto\sqrt{H}$, has been observed in YBCO and
LSCO~\cite{Hussey02,Wen04}. These have been taken as the bulk
evidence for the $d$-wave symmetry of the superconducting gap in
high-$T_c$ cuprates. Figure~\ref{fig.1} shows the $H$ dependence of
the $\gamma(H)$ for all doping samples. It can be seen the data are
well described by $\gamma(H)=A\sqrt{H}$ (shown as the solid curves)
with $A$ a doping-dependent constant. The numerical values of the
prefactor $A$ are listed in table~\ref{tab.1}. For completeness, the
$A$ for $x=0.19$ reported by Nohara \emph{et al.} is also
included~\cite{Nohara00}. The above result suggests that the
$d$-wave symmetry of the superconducting gap dominates the whole
phase diagram of LSCO, which is consistent with the result of the
recent phase-sensitive measurements~\cite{Tsuei04}. Furthermore, we
have shown that the field-induced SH is inversely proportional to
the nodal gap slope $v_\Delta$ (and the gap maximum $\Delta_0$),
that is, $A\propto1/v_\Delta\propto1/\Delta_0$~\cite{Wen05}. From
figure \ref{fig.1} and table \ref{tab.1} we can see the $A$
essentially decreases with decreasing doping. Thus this suggests
that the $\Delta_0$ becomes larger towards
underdoping~\cite{Wang07}. Actually it has been shown that in the
underdoped region, the $\Delta_0$ quantitatively tracks the
pseudogap in the normal state~\cite{Wen05}.

As the $H$ keeps rising, the sample would inevitably undergo a phase
transition into the normal state from the mixed state. In SH, this
means the field-induced DOS increases and eventually the normal
state electronic DOS is recovered. In other words, the $\gamma(H)$
rises and saturates to the normal state electronic SH coefficient
$\gamma_N$ at $H=H_{c2}$. Specifically, theory shows that for a
$d$-wave superconductor the relation between the above two
quantities can be written as
\begin{equation}
\label{eq.1}
\frac{\gamma(H)}{\gamma_N}=\sqrt{\frac{8}{\pi}}a\sqrt{\frac{H}{H_{c2}}},
\end{equation}
where $a$ is a constant depending only on the vortex lattice
geometry (=0.465 for a triangular vortex
lattice)~\cite{Volovik93,Kubert98}.

\begin{figure}
\onefigure{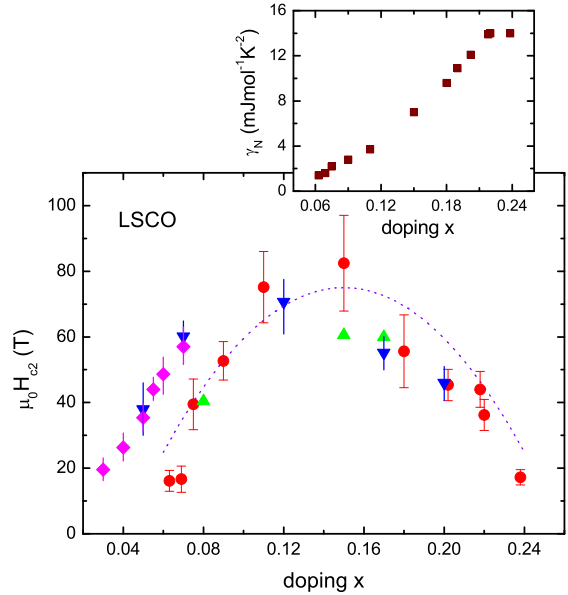} \caption{Doping dependence of the upper
critical field $H_{c2}(0)$ for LSCO obtained from SH (circles). The
error bars are estimated from the uncertainties of the $A$ values.
$H_{c2}(0)$ forms a ``dome" shape similar to the superconducting
transition temperature $T_c$. For comparison, values of $H_{c2}(0)$
for LSCO determined from resistive transport (up triangles,
Ref.~\cite{Ando99}), Nernst effect (down triangles,
Ref.~\cite{Wang06}) and torque magnetometry (diamonds,
Ref.~\cite{Li07}) are shown together. The dotted curve is a guide to
the eye. Inset: Coefficient of the normal-state electronic SH
$\gamma_N$ at $T=0$~K (Ref.~\cite{Matsuzaki04}). They are cited for
the evaluation of the $H_{c2}(0)$ shown in the main panel of the
figure. See text for details.} \label{fig.2}
\end{figure}

Combining eq.~(\ref{eq.1}) with the experimental result
$\gamma(H)=A\sqrt{H}$, we get $H_{c2}=8a^2\gamma_N^2/\pi A^2$. This
indicates that we may evaluate the $H_{c2}(0)$ from low-temperature
SH provided we could also access the $\gamma_N$ of the sample at
$T\rightarrow0$~K. Note that due to the large $H_{c2}(0)$ needed to
suppress the superconductivity, the $\gamma_N$ at $T\rightarrow0$~K
is hardly to be directly measured in SH for high-$T_c$ cuprates. One
alternative way is to investigate the superconducting transition of
the sample and extrapolate the $\gamma_N$ above $T_c$ to $T=0$~K
based on the entropy conservation. In principle, we should do this
performance in our own measurements. However, the uncertainty
associated with the separation of the electronic SH from the phonon
SH at elevated $T$ makes our relaxation technique not well suitable
for such a purpose. In this respect the community agrees that the
differential calorimetry may give more accurate
results~\cite{Fisher07}. Hence we instead look for the existing data
in the literature. Actually, Matsuzaki \emph{et al.} have recently
estimated the $\gamma_N$ at $T\rightarrow0$~K of LSCO across the
entire phase diagram in a systematic differential calorimetry
study~\cite{Matsuzaki04}. The doping dependence of the determined
$\gamma_N$ is reproduced in the inset of fig.~\ref{fig.2}. A roughly
linear increase of the $\gamma_N$ with increasing doping is
established up to $x\simeq0.2$. It is interesting to notice that the
electronic DOS at the Fermi level in LSCO revealed by the
angle-integrated photoemission spectroscopy (AIPES) follows
essentially the same doping evolution~\cite{Ino98}.

For underdoped cuprates, we note that the above result implies there
exists finite DOS in the ground state when superconductivity is
suppressed at $T=0$~K. Linked with the findings from the
angle-resolved photoemission spectroscopy (ARPES), it is natural to
speculate that this DOS resides on the Fermi arcs near ($\pi/2$,
$\pi/2$) nodal points for underdoped cuprates. ARPES has revealed
that at $T_c$ the Fermi surface is truncated forming arcs near the
nodal regions in the pseudogap phase~\cite{Norman98,Yoshida03}. It
is further found that with increasing doping the length of the arc
increases approximately linearly and at high doping level the arcs
eventually connect with each other near ($\pi$, 0) forming a large
Fermi surface~\cite{Yoshida06}. Therefore, to reconcile with the SH
it is not unreasonable to expect that for underdoped cuprates this
Fermi arc state persists to $T=0$~K as the ground state if the
superconductivity was totally suppressed, which results in the
finite DOS as shown in the SH~\cite{Wen07}. Note that this notion
has been supported by the independent nuclear magnetic resonance
(NMR) study in
Bi$_2$Sr$_{2-x}$La$_x$CuO$_{6+\delta}$~\cite{Zheng05}.

With the knowledge of the $A$ and $\gamma_N$, we have derived the
$H_{c2}(0)$ for LSCO according to the above expression, and plotted
the result in the main panel of fig.~\ref{fig.2}. The numerical
values are also presented in table~\ref{tab.1}. The determination of
$H_{c2}(0)$ in LSCO from specific heat allows us to see its general
behavior in a wide doping regime. As shown in fig.~\ref{fig.2},
towards either underdoping or overdoping, the $H_{c2}(0)$ falls from
its maximum value at optimal doping concomitant with the $T_c$,
forming a similar ``dome" shape as the $T_c$.

We have drawn the location of the $H_{c2}(0)$ in the $H$-$p$ diagram
for LSCO from SH. To gain more insight into the experimental result,
let us compare our $H_{c2}(0)$ with that determined by other
methods. Ando \emph{et al.} have reported the $c$ axis (interlayer)
resistivity in intense magnetic field $H\parallel c$ for
LSCO~\cite{Ando99}. Choosing 90\% of the normal-state resistivity as
criterion, the $H_{c2}$ was determined at low $T$ and is plotted as
the up triangles in fig.~\ref{fig.2}. We can see it shows good
consistency with our result. Note that a similar ``dome" shape of
the $H_{c2}(0)$ in Bi$_2$Sr$_2$CaCu$_2$O$_{8+y}$~(Bi2212) was also
obtained from the high-field interlayer resistive
transport~\cite{Krusin-Elbaum04}. In fig.~\ref{fig.2}, the down
triangles and diamonds represent the $H_{c2}$ for LSCO mapped in
Nernst effect and torque magnetometry measurements,
respectively~\cite{Wang06,Li07}. By extrapolating the high-field
Nernst or diamagnetic signal to zero, the scale of $H_{c2}$ has been
determined. It is shown that, aside from very underdoped region, the
$H_{c2}$ obtained in Nernst effect and in SH agree reasonably with
each other.

The quantitative agreement of the $H_{c2}(0)$ inferred from
different methods supports the validity of our estimation of the
$H_{c2}(0)$ from SH. Now we examine the implications of the
experimental findings. From above we have shown $A\propto1/\Delta_0$
and $H_{c2}\propto(\gamma_N/A)^2$, thus we obtain
$H_{c2}\propto(\gamma_N\Delta_0)^2$. This indicates that the
$H_{c2}(0)$ of the sample is governed by both the superconducting
pairing strength and the DOS contributing to the superconducting
condensation~\cite{Wen07}.  In the overdoped region, as shown in the
inset of fig.~\ref{fig.2}, the $\gamma_N$ increases slightly with
increasing doping and finally becomes almost a constant. In spite of
this, the $H_{c2}(0)$ is found to drop down monotonically as doping
increases from the optimal doped point. This means that it is the
reduction of the $\Delta_0$ that dominates the behavior of
$H_{c2}(0)$ in this region. In other words, the decrease of
$H_{c2}(0)$ with overdoping should originate mainly from the
reduction of the pairing strength.

Let us turn to the underdoped region, where the situation seems
different. The SH measured in underdoped region has revealed that
the $\Delta_0$ continues growing in the underdoped
region~\cite{Wen05}, which is also found by the thermal
conductivity~\cite{Takeya02,Sutherland03}. In analogy with the
overdoped side, one may expect the $H_{c2}(0)$ would keep rising
with underdoping provided that the pairing strength is still the
decisive factor to determine the $H_{c2}(0)$. Clearly this is at
odds with the present experimental result which shows that the
$H_{c2}(0)$ declines as doping reduces from the optimal doped point.
Therefore, contrary to the overdoped region, this indicates that in
the underdoped region it is the fall of the $\gamma_N$ that
overwhelms the rise of the $\Delta_0$ and leads to the dropping down
of the $H_{c2}(0)$ as the doping decreases. In the above we have
argued that in the underdoped region the $\gamma_N$ in SH
corresponds to the DOS on the Fermi arcs remaining in the pseudogap
phase. Thus we can say for underdoped cuprates the $H_{c2}(0)$
represents the field scale necessary to recover the DOS on Fermi
arcs from the superconducting state. Since the Fermi arcs shrink,
that is, the DOS available to the superconducting condensation
reduces towards underdoping, the $H_{c2}(0)$ naturally decreases
concomitantly.

The above analysis actually indicates that in the underdoped region
the coherent superconductivity below $T_c$ may be triggered by the
pairing of the carriers on nodal Fermi arcs. Since this pair process
would be associated with the formation of an energy gap near the
nodal region, it means that besides the pseudogap or the $\Delta_0$,
there is an effective superconducting energy scale, denoted as
$\Delta_{\mathrm{eff}}$, for underdoped cuprates. With decreasing
doping, though the $\Delta_0$ increases, the $\Delta_{\mathrm{eff}}$
actually decreases since it represents the maximum gap on nodal
Fermi arcs while the length of the Fermi arc reduces towards
underdoping. For underdoped cuprates, the $T_c$ or $H_{c2}(0)$ is
just the temperature or field scale necessary to close this
$\Delta_{\mathrm{eff}}$ in a BCS-like fashion,
respectively~\cite{Wen98}. As one increases the $T$ from below to
$T=T_c$ or applies the field $H=H_{c2}(0)$, the
$\Delta_{\mathrm{eff}}$ closes and the carriers on nodal Fermi arcs
are depaired, and therefore the coherent superconductivity is
destroyed with the appearance of nodal Fermi arcs while the
$\Delta_0$ near ($\pi$, 0) could remain unchanged. Note that this
suggestion seems to be consistent with the very recent ARPES
experiment which observed that a second energy gap opens at $T_c$
and has a BCS-like temperature dependence in underdoped
Bi2212~\cite{Lee07}. In contrast, for overdoped cuprates, however,
the situation is much simpler. With $H=H_{c2}(0)$, the
superconductivity disappears with the vanishing of the $\Delta_0$
and the recovery of the DOS on the large Fermi surface.

In the very underdoped region, it is worth noting that the
$H_{c2}(0)$ determined in Nernst effect and torque magnetometry
seems to be larger than that determined in the present SH. This
suggests that the $H_{c2}(0)$ probed by both methods may be
different in this very region. We believe that it may originate from
the strong superconducting fluctuations for underdoped
cuprates~\cite{Emery95}. In SH, since the $H_{c2}$ marks the
disappearance of the coherent superconductivity and the recovery of
the normal state electronic DOS near nodal regions, it vanishes as
the $T_c$ is approached from the low temperature. However, due to
sensitive to short-lived vortices, the Nernst effect or torque
magnetometry may detect the fluctuating superconducting correlations
above $T_c$ and thus give a higher $H_{c2}(0)$ which remains a
finite value at $T_c$. Note that a recent scanning tunnelling
microscopy (STM) experiment in Bi2212 reported a nucleation of
pairing gaps in nanoscale regions above T$_c$, which was considered
as a microscopic basis for the above fluctuating superconducting
response~\cite{Gomes07}. Under this circumstance, we view the
$H_{c2}(0)$ defined in our SH as the field scale to destroy the
phase coherence of the superconductivity in the very underdoped
region and thus have the suggestion that the pairing of the carriers
on nodal Fermi arcs is crucial to and occurs simultaneously with the
onset of the phase coherence. While at the same region the
$H_{c2}(0)$ defined in Nernst effect or torque magnetometry may mark
a higher field scale to destroy the phase-disordered condensate.
Note that this is in particular indicated by the fact that, as shown
in fig. \ref{fig.2}, the $H_{c2}(0)$ probed in torque magnetometry
varies continuously with $x$ down to $x=0.03$ while the coherent
superconductivity only appears down to $x\approx0.055$~\cite{Li07}.
On the other hand, it is interesting to note that for underdoped
LSCO the $H_{c2}(0)$ probed in SH is actually roughly consistent in
value with the field revealed in torque magnetometry  which marks
the melt of the vortex solid at $T\rightarrow0$ K and falls to zero
as $x\rightarrow0.055$~\cite{Li07}. In the end, We mention that the
above speculation is also backed by the observation that in the
overdoped region where fluctuations are most weak or absent the
$H_{c2}(0)$ determined in different methods agree with each other.

\begin{figure}
\onefigure{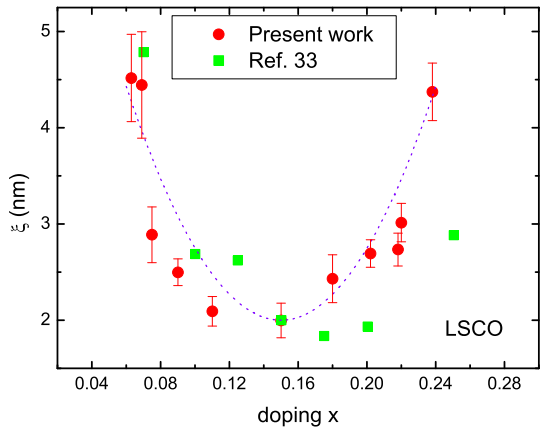} \caption{Doping dependence of the coherence
length $\xi$ for LSCO derived from the $H_{c2}$ with
$H_{c2}=\Phi_0/(2\pi\xi^2)$ (circles). For comparison, the coherence
length $\xi$ obtained from magnetization measurements for LSCO
(squares, Ref.~\cite{Wen03}) is also plotted. The dotted curve is a
guide to the eye. The results indicate a growing coherence length
$\xi$ from optimal doping point towards either underdoping or
overdoping.} \label{fig.3}
\end{figure}

Finally, from $H_{c2}=\Phi_0/(2\pi\xi^2)$, we can obtain the doping
dependence of the coherence length $\xi$. Figure~\ref{fig.3} shows
the calculated $\xi$ from the $H_{c2}(0)$ (circles). It is shown
that from the optimal doping point the $\xi$ grows towards
underdoping or overdoping. Previously the doping dependence of
$\xi$, regarded as the size of the vortex core, had been drawn from
the systematic magnetization measurements in LSCO thin
films~\cite{Wen03}, which is also plotted in fig. \ref{fig.3}
(squares). It can be seen that both experiments show a reasonable
consistency, which assures again the reliability of obtaining
$H_{c2}$ from low-temperature SH. Moreover, it should be noted that
an increase of the $\xi$ with underdoping from the optimal doping
point has also been suggested by the fluctuation
magneto-conductivity~\cite{Ando02} and the reversible
magnetization~\cite{Gao06} measurements in
YBa$_2$Cu$_3$O$_{7-\delta}$ (YBCO). Furthermore, it could be
instructive to compare $\xi$ with the the Pippard coherence length
$\xi_p=\alpha\hbar v_F/k_BT_c$, where $v_F$ is the Fermi velocity
and $\alpha$ a numerical constant of order unity~\cite{Tinkham96}.
As $v_F$ is nearly doping independent~\cite{Zhou03}, $\xi_p\propto
1/T_c$ and thus the variation of $\xi_p$ with doping is
qualitatively the same as that of $\xi$ since both $T_c$ and
$H_{c2}$ form a ``dome" shape with doping. In terms of $\Delta_0$,
on the other hand, $\xi_p$ can be also expressed as $\xi_p=\hbar
v_F/\beta\Delta_0$ with $\beta$ another numerical constant. Note
that there seems to be inconsistency between the above two
expressions for $\xi_p$ for underdoped region since $T_c$ and
$\Delta_0$ have opposite doping dependence. Actually this seeming
contradiction could be naturally resolved by substituting $\Delta_0$
with $\Delta_{\mathrm{eff}}$ in the expression of $\xi_p$, that is,
$\xi_p=\hbar v_F/\beta\Delta_{\mathrm{eff}}$ for underdoped
cuprates. This further suggests that there may be an effective
superconducting energy scale determining the coherent
superconductivity in the underdoped region.

\section{Conclusion}

In summary, we have analyzed the low-temperature SH in LSCO
throughout the whole superconducting dome to evaluate the
zero-temperature $H_{c2}$ which is defined as the field scale
necessary to remove the coherent superconductivity and recover the
normal state electronic DOS. It is found that the doping dependence
of $H_{c2}(0)$ essentially follows that of $T_c$. In the underdoped
region, the decrease of $H_{c2}(0)$ concomitant with $T_c$ suggests
that the coherent pairing of the carriers on nodal Fermi arcs plays
important role in establishing the high-temperature
superconductivity.

\acknowledgments

This work was supported by the National Science Foundation of China,
the Ministry of Science and Technology of China (973 Projects No.
2006CB601000, No. 2006CB921802), and Chinese Academy of Sciences
(Project ITSNEM).

\end{document}